\documentclass[final,3p,times,twocolumn]{elsarticle}
 \biboptions{comma,sort&compress}
\usepackage{here}
 \usepackage{graphicx}
  \usepackage{epsfig}
\def\nin{\noindent}
\def\beq{\begin{equation}}
\def\eeq{\end{equation}}
\def\bea{\begin{eqnarray}}
\def\eea{\end{eqnarray}}

\begin{document}

\begin{frontmatter}



\title{Closure testing the NNPDF3.0 methodology}

 \author{Christopher S.\ Deans, on behalf of the NNPDF collaboration}
  \address{The Higgs Centre for Theoretical Physics, University of Edinburgh,\\
JCMB, KB, Mayfield Rd, Edinburgh EH9 3JZ, Scotland}

\begin{abstract}
\noindent
A thorough understanding of the issues surrounding the determination of parton distributions is crucial due to their importance to calculations of LHC observables. 
However, it is still not fully understood how much of an impact methodological bias has on PDF fits.
Closure tests, where a fit is performed to pseudo-data generated using an existing PDF set, provide a way of directly investigating whether current PDF fitting methodologies are successful.
Here, we present a sample of results from closure tests applying the NNPDF methodology to data created using a variety of different PDF sets.
The results validate our methodology by showing that the initial PDFs can be reproduced within uncertainties.
We also briefly discuss our latest PDF determination, NNPDF3.0, which has been developed making extensive use of the closure test technique.

\end{abstract}

\end{frontmatter}

% \linenumbers

\section{Introduction}
\label{sec-int}

\nin
Parton distribution functions (PDFs) provide a description of the internal structure of the proton and are an important input into theoretical calculations of hadronic processes.
PDFs and their associated uncertainties are a major contribution to the uncertainty on many LHC measurements, and this will only increase as data as well as perturbative calculations become more precise. 
It is therefore very important that the there is a good understanding of how PDF uncertainties are obtained, so that we can be confident that they are correctly determined.
However, comparison of results from different PDF determinations~\cite{BENCH} demonstrate that, while there is general agreement between the three global PDF sets, there are still some unresolved discrepancies.
One common idea is that these differences are at least partly caused by differences in methodology between the different determinations.

%%%%%%%%%

The NNPDF approach to determining PDFs is aimed at reducing bias stemming from the methodology and provide robust PDF uncertainties~\cite{NN23}\cite{NN20}\cite{NN10}.
Our PDFs are parametrized by feed forward neural networks, which are extremely flexible.
To characterize the PDF uncertainties we produce a set of Monte Carlo replica PDFs, from which central values, standard deviations and any other statistical measures can be easily obtained.

%%%%%%%%%

In addition to these features, for our latest release we have made use of closure tests to verify the success of our methodology.
In a closure test, a fit is performed to pseudo-data generated using an existing PDF set.
The results of the closure test fit can then be compared directly to the `correct' answer.
In this proceedings I will first briefly describe the latest NNPDF determination, NNPDF3.0, and then present some results from closure tests of the NNPDF methodology.  

%%%%%%%%%%%%%%%%%%%%%%%
\begin{figure*}[t]
\centering
\includegraphics[width=0.46\textwidth]{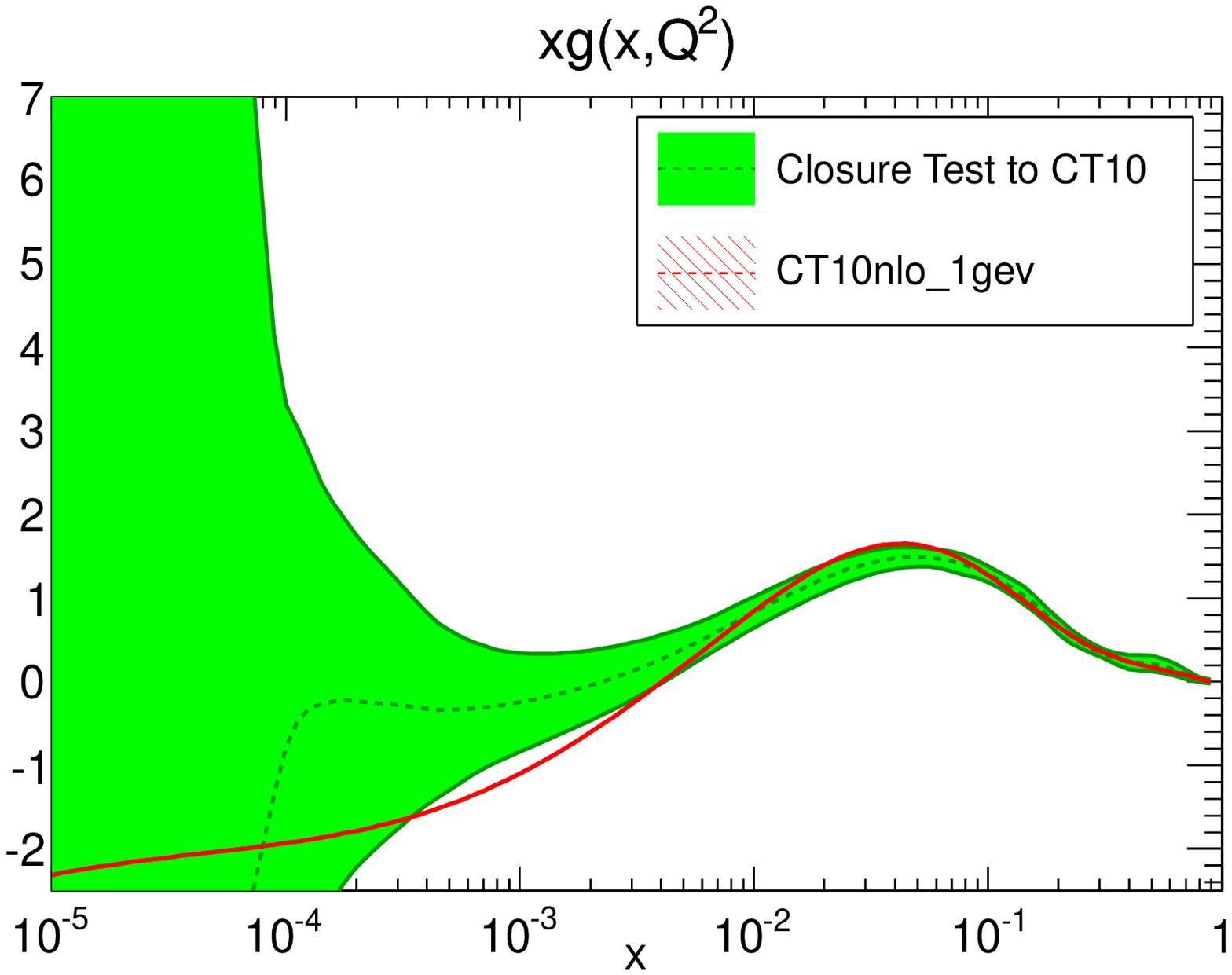}
\includegraphics[width=0.46\textwidth]{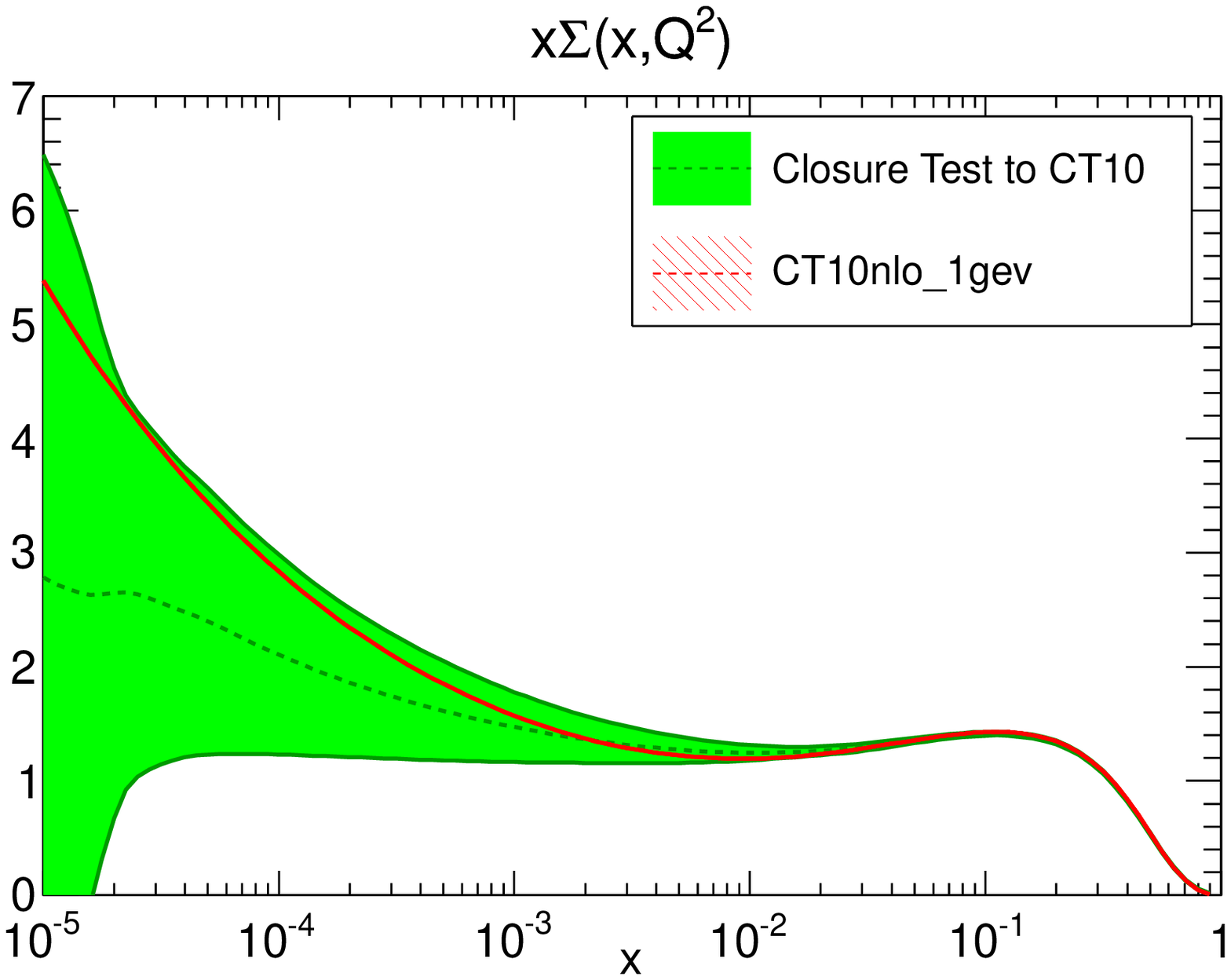}
%{\epsfig{figure=mpsi2mc.eps,height=70mm}}
\caption{\scriptsize Gluon and Singlet PDFs for central value of CT10 NLO (red) and closure test fit to pseudo-data generated from CT10 (green). The PDFs are plotted at $Q^2 = 1$ $\mathrm{GeV}^2$ which is the scale at which the PDFs are parametrized in the fit. To evolve CT10 down to this scale we used APFEL~\cite{APFEL}.}
\label{figCT} 
\end{figure*} 
%%%%%%%%%%%%%%%%

\section{NNPDF3.0}
\label{sec-nn30}

NNPDF3.0 is the latest PDF determination from the NNPDF collaboration~\cite{NN30}.
It combines the inclusion of a wide range of new data, much of it included for the first time in a global PDF analysis, along with several theory improvements and a completely renovated and validated fitting methodology.

%%%%%%%%

NNPDF3.0 uses over 1000 new datapoints, which combines with previously included data for a total of 4276 points.
Amongst the new data are HERA-II cross section data from both H1 and ZEUS, as well as combined charm production data.
We have also added a substantial amount of new LHC data.
From ATLAS there are $W$ $p_T$ data, high mass Drell-Yan data, and 2.76 TeV inclusive jet data fully correlated with the already included 7 TeV data.
From CMS we now include 7 TeV jet data from 2011, $W$ muon asymmetry, double differential Drell-Yan and $W+$charm data.
We also include some $Z$ production data from LHCb.

%%%%%%%%

In addition to the expanded dataset, NNPDF3.0 also boasts a significantly improved methodology.
A substantial amount of work has been done refining and tuning the genetic algorithm used to fit the neural networks.
The range of positivity constraints used during the fit has also been expanded, both in terms of types of observables used and the kinematic range covered.
We have also improved the stopping procedure used to prevent over-learning.
There is also significant improvements in the theory caluclations used in the fit, both at NLO with the development of fast interfaces like APPLGRID~\cite{APPL} and aMCfast~\cite{AMCFAST}, and at NNLO with new calculation for top data~\cite{TOP} and jet data~\cite{JET} amonst others.

The central NNPDF3.0 PDF sets are already available on LHAPDF, and a paper describing the analysis in detail is due to be released soon.

%%%%%%%%%%%%
\section{Closure Tests}
\label{sec-ct}
\nin

%%%%%%%%

In a normal PDF fit to real experimental data, the underlying law which we are trying to estimate is unknown.
This makes it difficult to evaluate how well a fitting methodology can reproduce the `correct' answer and to ascertain whether there are sources of bias.
In addition, all fits are to the same particular data, so it is unclear whether an improvement in the quality of the fit to these data represents an actual improvement or is due to over-learning, where the noise in the data is fitted along with the underlying pattern.
%
%The possibility of inconsistencies in the data or inadequate theoretical descriptions further complicate an analysis of the methodology with real data.
%
Closure tests provide a way to evaluate the success of a fitting methodology which avoids these issues.
By fitting pseudo-data generated from an existing PDF set we can compare directly the results of the fit with the known underlying law.

%%%%%%%%

The closure testing procedure is as follows.
First we take the real data and replace the datapoints with the central values calculated using a given PDF set.
Then the experimental uncertainties are used to generate artificial fluctuations in these new central values.
This results in a set of pseudo-data which is perfectly consistent, both to it's quoted uncertainties and to the theory used in the fit.
This pseudo-data set can then be fit using the same NNPDF methodology as is used in fits to real data.
The result is a PDF fit for which the `true' result is known: the original PDF set used to generate the data.
Note that the PDF uncertainties on the closure test fit are still driven by the experimental uncertainties, at least in the data region, and are unrelated to the uncertainties on the generating PDFs, which do not enter into the process. 

%%%%%%%%

The option to perform a closure test has been implemented in the NNPDF code in a very flexible way.
Any PDF set can be used to generate the pseudo-data, with the only technical limitation being that it needs to be provided to the code in the LHAPDF format.
Multiple different pseudo-data sets can also be generated by changing the random seed.
It is also possible to generate data without random noise, providing an environment in which to test the fitting algorithm without the possibility of over-learning. 

%%%%%%%%

As part of the work leading up to the release of our latest PDF set, we have performed a large number of closure tests with a variety of different settings.
In the past we have also looked at closure tests to simple forms~\cite{NHCDPROC}, and there has also been work by MSTW on using closure tests to investigate the effects of inconsistent data~\cite{WATT}.
Here I will present a small portion of the results that we obtained, specifically looking at closure test fits to different PDF sets.
A more encompassing look at NNPDF3.0 closure tests will be available in our upcoming paper~\cite{NN30}.

%%%%%%%%
%%%%%%%%%%%%%%%%
\begin{figure*}[t]
\centering
\includegraphics[width=0.46\textwidth]{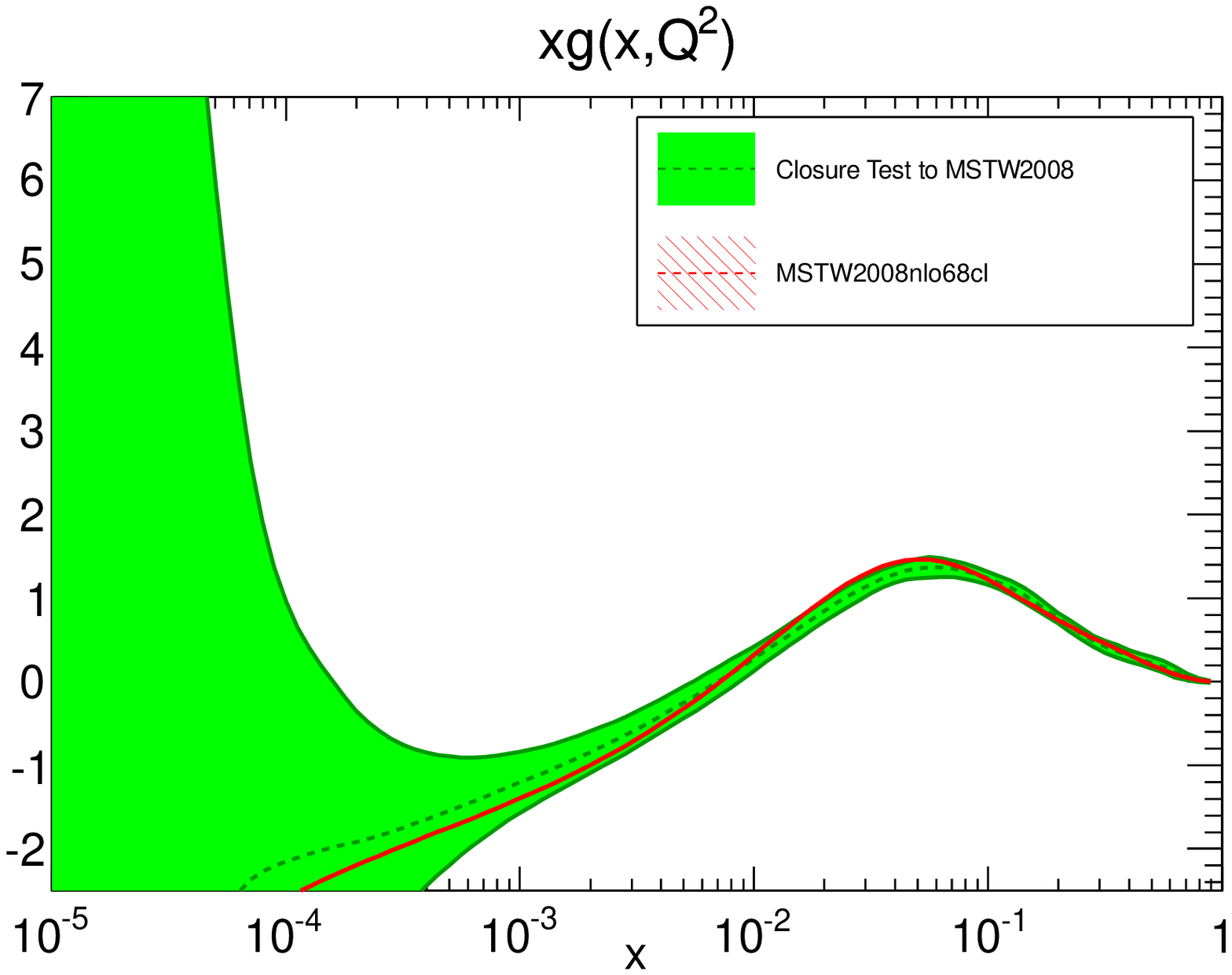}
\includegraphics[width=0.46\textwidth]{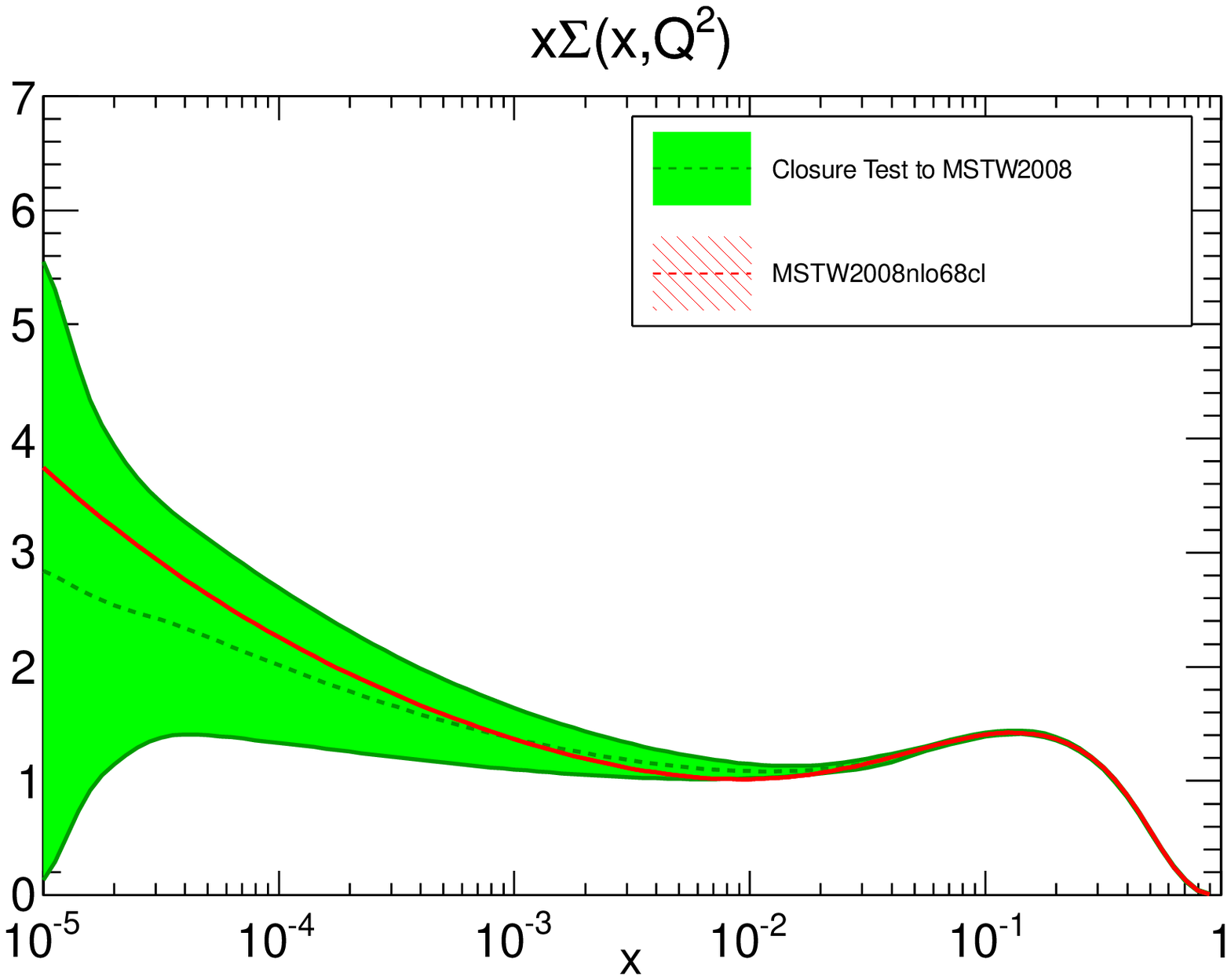}
%{\epsfig{figure=mpsi2mc.eps,height=70mm}}
\caption{\scriptsize Gluon and Singlet PDFs for central value of MSTW2008 NLO (red) and closure test fit to pseudodata generated from MSTW2008 (green). The PDFs are plotted at $Q^2 = 1$ $\mathrm{GeV}^2$.}
\label{figMSTW} 
\end{figure*} 
%%%%%%%%%%%%%%%%
\begin{figure*}[t]
\centering
\includegraphics[width=0.46\textwidth]{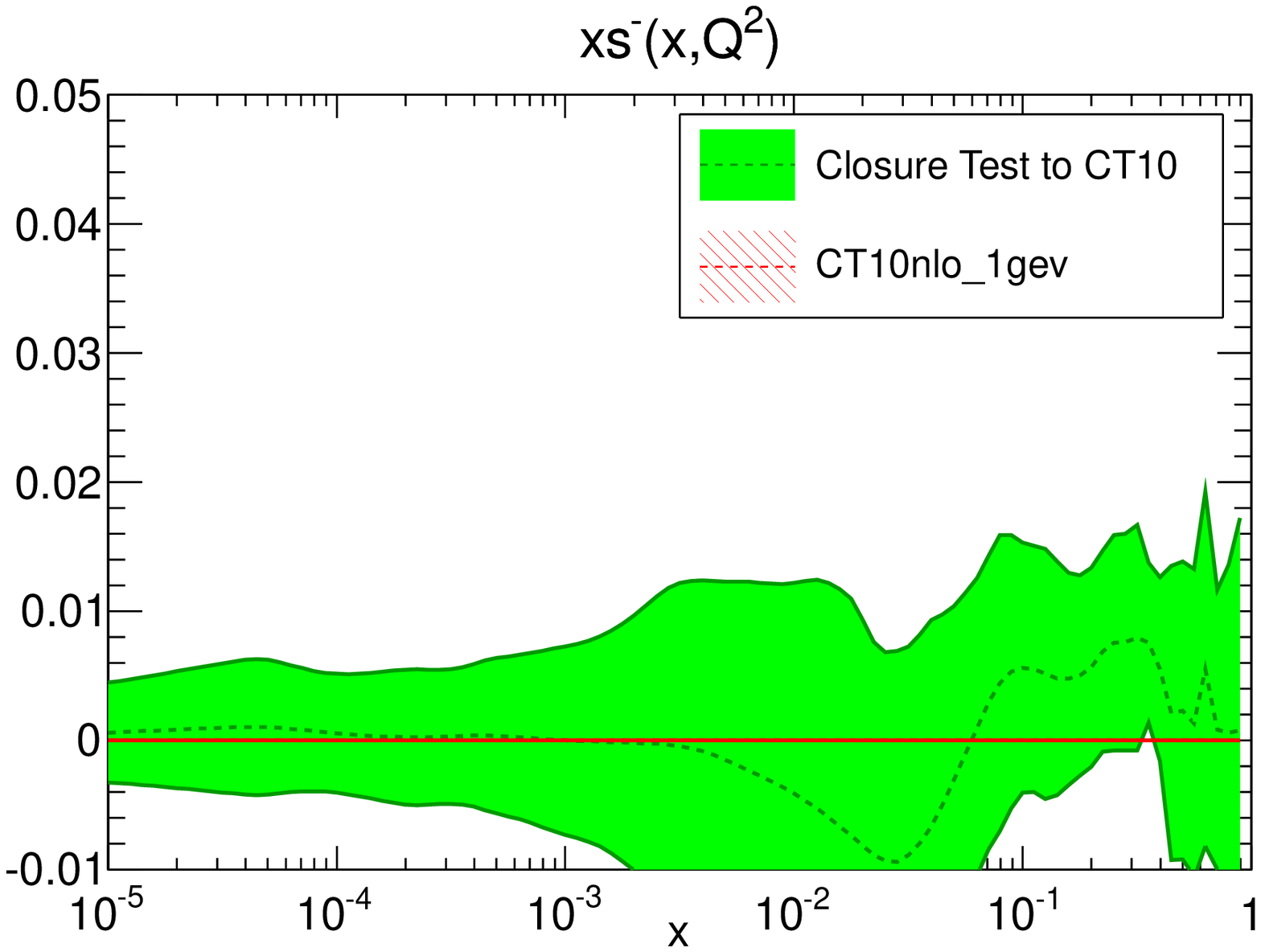}
\includegraphics[width=0.46\textwidth]{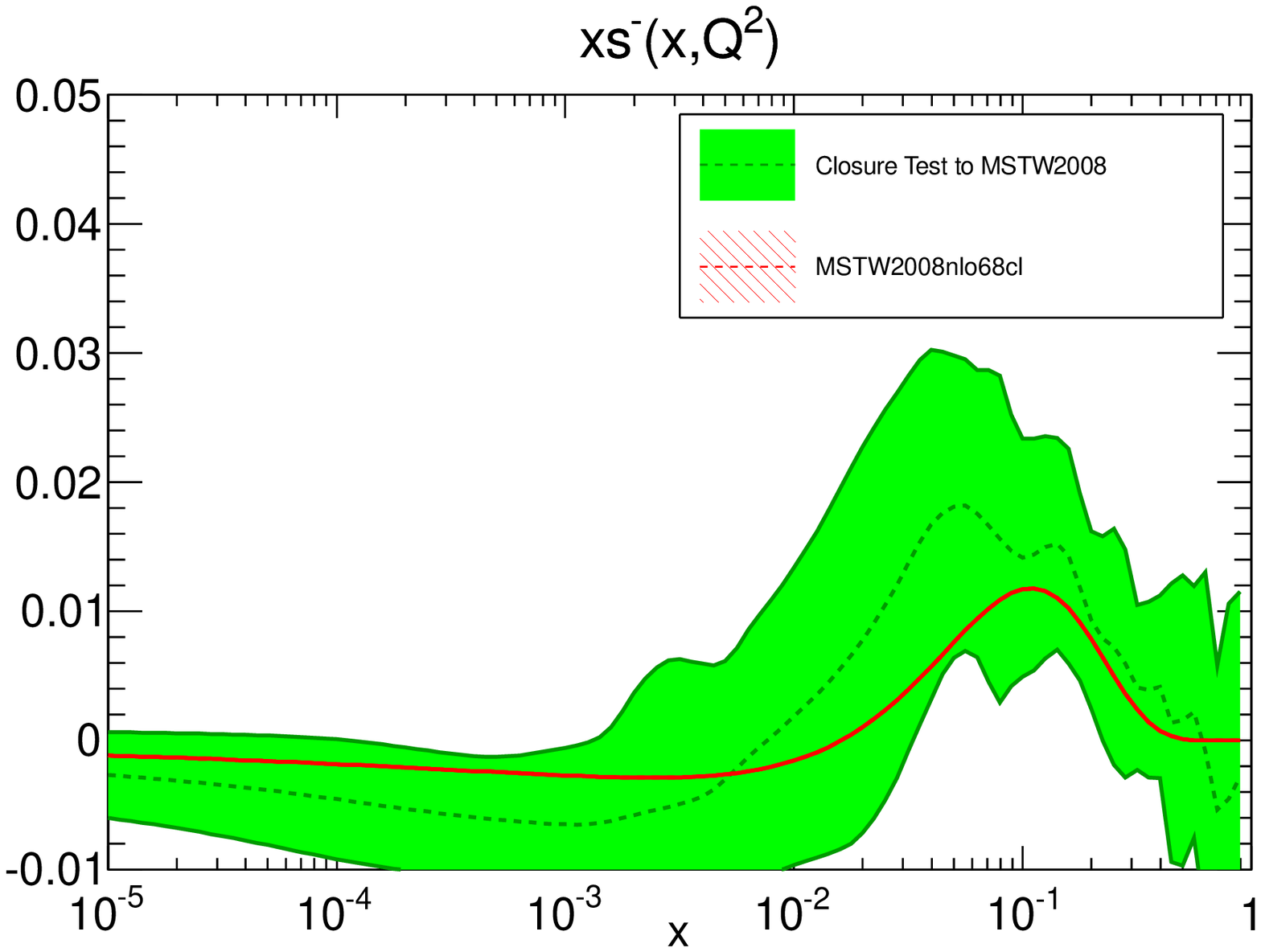}
%{\epsfig{figure=mpsi2mc.eps,height=70mm}}
\caption{\scriptsize Strange valence PDF for closure test fits (green) and the central PDF used to generate the closure test pseudodata (red). On the left is CT10 and on the right MSTW2008. PDFs are plotted at $Q^2 = 1$ $\mathrm{GeV}^2$.}
\label{figSM} 
\end{figure*}
%%%%%%%%%%%%%%%%

\section{Results of Closure Tests}
\label{sec-res}
\nin

A closure test fit was performed using the CT10 PDF set~\cite{CT10}.
The data was generated using the NLO PDF set and theory, though for closure tests the details of the theory are irrelevant as long as the same theory is used in both the data generation and fit.
The fit were performed using the full NNPDF3.0 methodology, except for the positivity constraints which were not applied as these were not completely satisfied by the input set.
%
%Additionally, it was necessary to evolve the CT10 PDFs down to $Q^2 = 1$ $\mathrm{GeV}^2$ as currently our fitting code can only calculate theoretical observables from PDFs at this scale. 

%%%%%%%%

Some of the results from these closure test are shown in Figure~\ref{figCT}. 
The plots compare the gluon and singlet obtained from the closure test fit in green to the central value of the CT10 PDF set in red.
The closure test fit is in good agreement with the original PDFs in both cases.
This level of constancy is also seen in the other PDFs fit in the closure test, which are not shown here.
We expect for a successful closure test the PDF uncertainties should contain the theory value about 68\% of the time, and we have checked explicitally that this is the case.

%%%%%%%%%%%%%%%%
\begin{figure*}[t]
\centering
\includegraphics[width=0.46\textwidth]{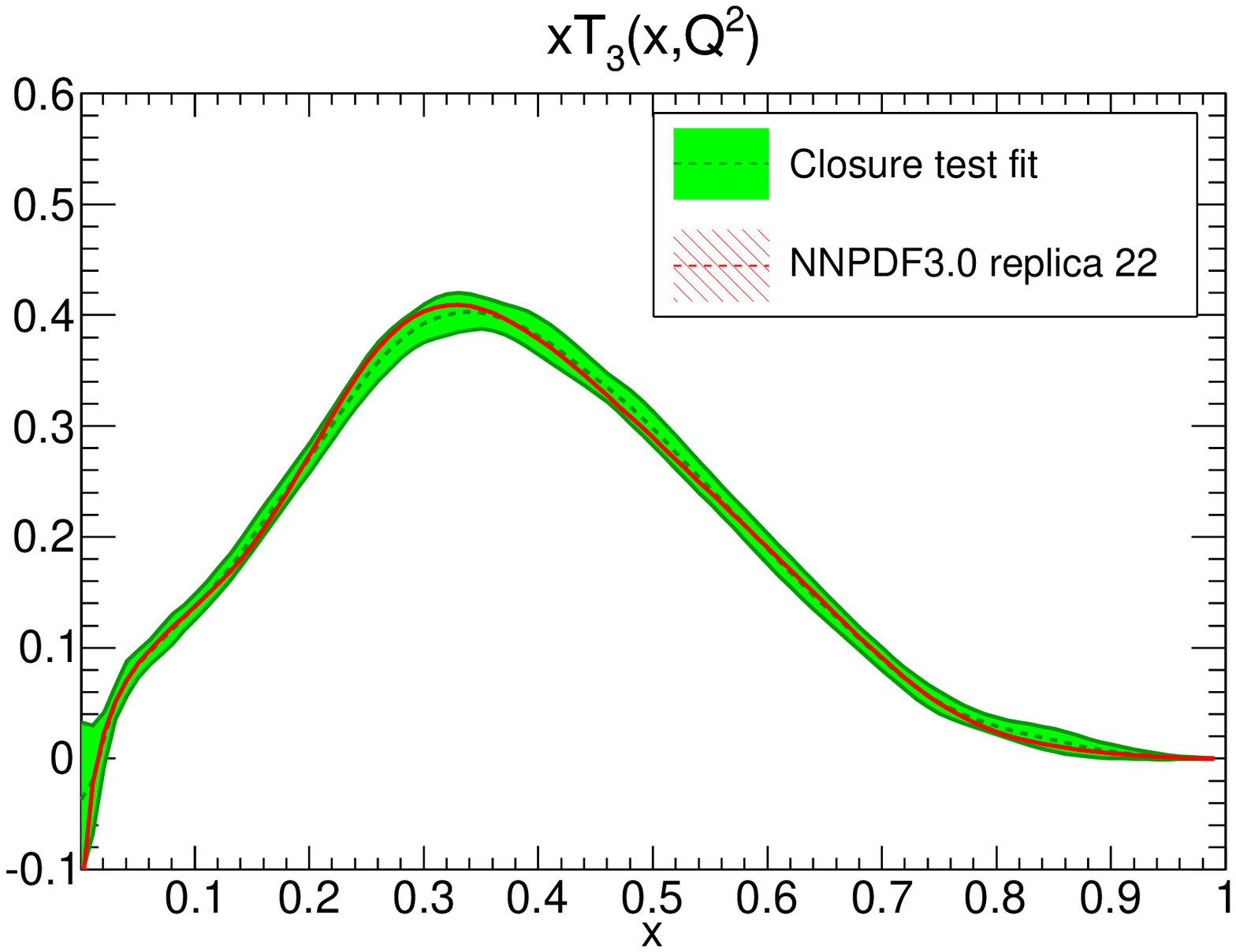}
\includegraphics[width=0.46\textwidth]{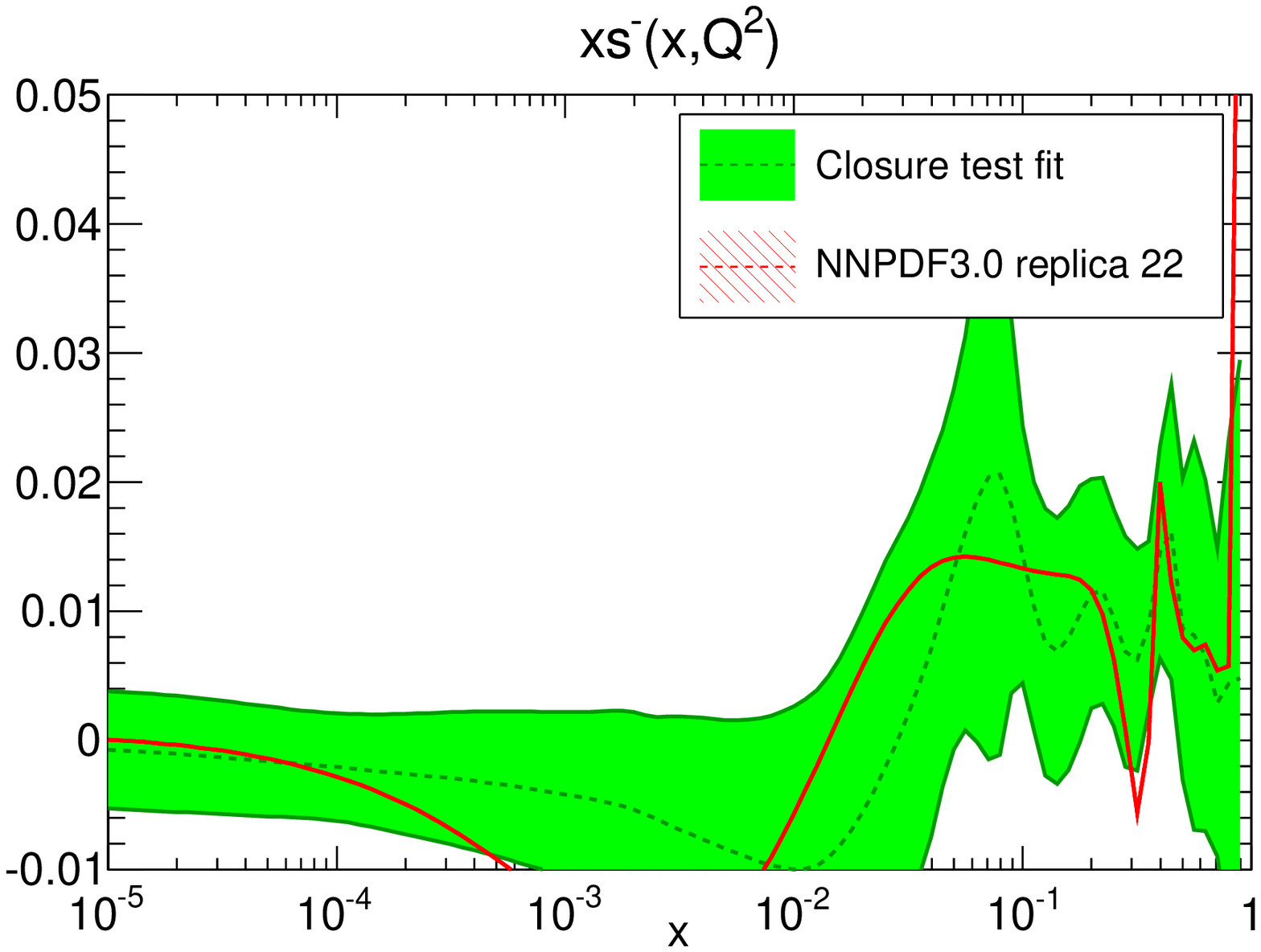}
%{\epsfig{figure=mpsi2mc.eps,height=70mm}}
\caption{\scriptsize Triplet and strange valence PDFs for replica 22 of NNPDF3.0 NLO (red) and closure test fit to pseudo-data generated from that replica (green). The triplet is plotted on a linear scale in $x$, while the strange valence is plotted on a log scale. The PDFs are plotted at $Q^2 = 1$ $\mathrm{GeV}^2$.}
\label{figNN30} 
\end{figure*} 
%%%%%%%%%%%%%%%%

Figure~\ref{figMSTW} shows results from a similar closure test fit, but this time performed using pseudo-data generated using the MSTW2008 NLO PDF set~\cite{MSTW}.
Again the singlet and gluon PDFs are reconstructed successfully at the one sigma level.

For the singlet and gluon, both MSTW2008 and CT10 have similar shaped PDFs, so results of the two closure test fits are also very similar.
However we can look at example where this is not the case.
Figure~\ref{figSM} shows the strange valence distributions ($s - \bar{s}$) for the central value of CT10 and MSTW2008 in red and from the corresponding closure test in green.
This PDF combination is constrained to be zero for all $x$ in CT10, but is allowed to have more structure in MSTW2008.
In both cases the closure test fit correctly models the underlying distribution, though with larger uncertainties than for the gluon and singlet due to the poorer data resolution for this combination.
This shows the flexibility of the NNPDF parametrization in that it can successfully model either shape.
It also demonstrates the ability of closure tests to probe how much the PDFs are constrained by data.
Here, the fact that the closure test fit can correctly determine whether there is a non-zero strange valence distribution indicates that there enough relevant data to do this also in fits to real data.

%%%%%%%%%%

While both of these closure tests demonstrate the ability of the NNPDF methodology to successfully fit PDFs with a smooth shape, it is also possible to perform a closure test using a more exotic underlying theory.
As an example of this, Figure~\ref{figNN30} shows some of the results from a closure test fit to a single Monte Carlo replica from the NNPDF3.0 NLO PDF set.
On the left is shown the large-$x$ triplet ($u+\bar{u}-d-\bar{d}$) and on the right the strange valence distribution, with again the closure test fit compared to the central value of the PDFs used to create the pseudo-data.
The triplet is extremely well reproduced in the closure test, despite the various wiggles in the replica PDF.
The very oddly shaped strange valence is less closely matched but the closure test fit is still consistent within uncertainties.

\section{Outlook}
\label{sec-out}
\nin

These closure test results demonstrate that the NNPDF methodology is capable of correctly reproducing a known underlying law without any large signs of bias.
%
%This is seen to be true for tests using multiple different initial PDF sets, including very unrealistic ones.
%
This therefore gives us confidence in the results we obtain in fits to the real experimental data.
We can also see that the closure test technique itself is very powerful, both for validating our methodology and also for investigating the impact of data in the case where it is perfectly consistent with existing sets.
Closure tests are therefore an important tool for both current and future NNPDF analyses.



%% References


%\bibliographystyle{elsarticle-num}





% \end{thebibliography}

%%%%%%%%%%%%%%%%%%%%
%\vfill\eject

%%%%%%%%%%%%%%%

\end{document}